\documentstyle[preprint,eqsecnum,aps,epsfig,floats,tighten]{revtex}

\def \GeVmass {\:{\rm GeV}}
\def \TeV {\:{\rm TeV}}
\def\etal{{\sl et al.}}
\def\dzero{D\O\ }
\def\Journal#1#2#3#4{{#1} {\bf #2}, #3 (#4)}

\def\CPC{Comput. Phys. Commun.}

\def\NIM{Nucl. Instrum. Methods}

\def\NPBproc{Nucl. Phys. B (Proc. Suppl.)}
\def\PLB{Phys. Lett. B}

\def\PRL{Phys. Rev. Lett.}
\def\PRD{Phys. Rev. D}

\def\EPC{Eur. Phys. J. {\bf C}$\!$}

\begin{document}
\draft
\preprint{\sc{FERMILAB-Pub-00/046-E}}
\title{Spin Correlation in $t\bar{t}$ Production from $p\bar{p}$ Collisions
\\ at $\sqrt{s}=1.8$ TeV}

%
\author{                                                                      
B.~Abbott,$^{46}$                                                             
M.~Abolins,$^{43}$                                                            
V.~Abramov,$^{19}$                                                            
B.S.~Acharya,$^{13}$                                                          
D.L.~Adams,$^{53}$                                                            
M.~Adams,$^{30}$                                                              
V.~Akimov,$^{17}$                                                             
G.A.~Alves,$^{2}$                                                             
N.~Amos,$^{42}$                                                               
E.W.~Anderson,$^{35}$
% ...
\\                                                                            
\vskip 0.30cm                                                                 
\centerline{(D\O\ Collaboration)}                                             
\vskip 0.30cm                                                                 
}                                                                             
\address{                                                                     
\centerline{$^{1}$Universidad de Buenos Aires, Buenos Aires, Argentina}       
\centerline{$^{2}$LAFEX, Centro Brasileiro de Pesquisas F{\'\i}sicas,         
                  Rio de Janeiro, Brazil}                                     
\centerline{$^{3}$Universidade do Estado do Rio de Janeiro,                   
                  Rio de Janeiro, Brazil}                                     
\centerline{$^{4}$Institute of High Energy Physics, Beijing,                  
                  People's Republic of China}                                 
\centerline{$^{5}$Universidad de los Andes, Bogot\'{a}, Colombia}             
\centerline{$^{6}$Universidad San Francisco de Quito, Quito, Ecuador}         
\centerline{$^{7}$Institut des Sciences Nucl\'eaires, IN2P3-CNRS,             
                  Universite de Grenoble 1, Grenoble, France}                 
\centerline{$^{8}$CPPM, IN2P3-CNRS, Universit\'e de la M\'editerran\'ee,      
                  Marseille, France}                                          
\centerline{$^{9}$LPNHE, Universit\'es Paris VI and VII, IN2P3-CNRS,          
                  Paris, France}                                              
\centerline{$^{10}$DAPNIA/Service de Physique des Particules, CEA, Saclay,    
                  France}                                                     
\centerline{$^{11}$Panjab University, Chandigarh, India}                      
\centerline{$^{12}$Delhi University, Delhi, India}                            
\centerline{$^{13}$Tata Institute of Fundamental Research, Mumbai, India}     
\centerline{$^{14}$Seoul National University, Seoul, Korea}                   
\centerline{$^{15}$CINVESTAV, Mexico City, Mexico}                            
\centerline{$^{16}$Institute of Nuclear Physics, Krak\'ow, Poland}            
\centerline{$^{17}$Institute for Theoretical and Experimental Physics,        
                   Moscow, Russia}                                            
\centerline{$^{18}$Moscow State University, Moscow, Russia}                   
\centerline{$^{19}$Institute for High Energy Physics, Protvino, Russia}       
\centerline{$^{20}$Lancaster University, Lancaster, United Kingdom}           
\centerline{$^{21}$University of Arizona, Tucson, Arizona 85721}              
\centerline{$^{22}$Lawrence Berkeley National Laboratory and University of    
                   California, Berkeley, California 94720}                    
\centerline{$^{23}$University of California, Davis, California 95616}         
\centerline{$^{24}$California State University, Fresno, California 93740}     
\centerline{$^{25}$University of California, Irvine, California 92697}        
\centerline{$^{26}$University of California, Riverside, California 92521}     
\centerline{$^{27}$Florida State University, Tallahassee, Florida 32306}      
\centerline{$^{28}$University of Hawaii, Honolulu, Hawaii 96822}              
\centerline{$^{29}$Fermi National Accelerator Laboratory, Batavia,            
                   Illinois 60510}                                            
\centerline{$^{30}$University of Illinois at Chicago, Chicago,                
                   Illinois 60607}                                            
\centerline{$^{31}$Northern Illinois University, DeKalb, Illinois 60115}      
\centerline{$^{32}$Northwestern University, Evanston, Illinois 60208}         
\centerline{$^{33}$Indiana University, Bloomington, Indiana 47405}            
\centerline{$^{34}$University of Notre Dame, Notre Dame, Indiana 46556}       
\centerline{$^{35}$Iowa State University, Ames, Iowa 50011}                   
\centerline{$^{36}$University of Kansas, Lawrence, Kansas 66045}              
\centerline{$^{37}$Kansas State University, Manhattan, Kansas 66506}          
\centerline{$^{38}$Louisiana Tech University, Ruston, Louisiana 71272}        
\centerline{$^{39}$University of Maryland, College Park, Maryland 20742}      
\centerline{$^{40}$Boston University, Boston, Massachusetts 02215}            
\centerline{$^{41}$Northeastern University, Boston, Massachusetts 02115}      
\centerline{$^{42}$University of Michigan, Ann Arbor, Michigan 48109}         
\centerline{$^{43}$Michigan State University, East Lansing, Michigan 48824}   
\centerline{$^{44}$University of Nebraska, Lincoln, Nebraska 68588}           
\centerline{$^{45}$Columbia University, New York, New York 10027}             
\centerline{$^{46}$New York University, New York, New York 10003}             
\centerline{$^{47}$University of Rochester, Rochester, New York 14627}        
\centerline{$^{48}$State University of New York, Stony Brook,                 
                   New York 11794}                                            
\centerline{$^{49}$Brookhaven National Laboratory, Upton, New York 11973}     
\centerline{$^{50}$Langston University, Langston, Oklahoma 73050}             
\centerline{$^{51}$University of Oklahoma, Norman, Oklahoma 73019}            
\centerline{$^{52}$Brown University, Providence, Rhode Island 02912}          
\centerline{$^{53}$University of Texas, Arlington, Texas 76019}               
\centerline{$^{54}$Texas A\&M University, College Station, Texas 77843}       
\centerline{$^{55}$Rice University, Houston, Texas 77005}                     
\centerline{$^{56}$University of Washington, Seattle, Washington 98195}       
}                                                                             
\date{\today}
\maketitle

\newpage

\begin{abstract}
The \dzero collaboration has performed a study of spin
correlation in $t\bar{t}$ production 
for the process $t\bar{t} \rightarrow bW^+\bar{b}W^-$,
where the $W$ bosons decay to $e\nu$ or $\mu\nu$. A 
sample of six events was collected during an exposure of the \dzero detector
to an integrated luminosity of approximately 125 ${\rm pb}^{-1}$ of $\sqrt{s}=1.8\ \TeV$ 
$p\bar{p}$ collisions. The standard model (SM) predicts that
the short lifetime of the top quark ensures
the transmission of any spin information at production to
the $t\bar{t}$ decay products.
 The degree of spin correlation 
is characterized by a correlation coefficient $\kappa$. 
We find that $\kappa>-0.25$ at the 68\% confidence level, in 
agreement with the SM prediction of $\kappa=0.88$.
\end{abstract}

\newpage
\renewcommand{\theequation}{\arabic{equation}}
Pair production of top quarks has been observed \cite{top_papers} in $p\bar{p}$
collisions at $\sqrt{s}=1.8\ \TeV$ by both the CDF and \dzero collaborations,
and the mass and production cross section have been measured
in various channels \cite{top_recent,CDF}. The observed properties
agree well with predictions from the standard model (SM).

For a top quark mass of  $m_t=175$~GeV, the width of the top quark in the 
SM is $\Gamma_t=1.4$~GeV \cite{Jezabek} while the typical hadronization
scale is $\Lambda_{\rm QCD}\approx 0.22 \:\GeVmass$ \cite{PDG}.
The time scale
needed for depolarization of the top-quark spin is of the order 
$m_t/\Lambda_{\rm QCD}^2\gg 1/\Gamma_t$
\cite{Peskin}, implying that  
polarization information should be transmitted fully to the decay products
of the top quark.
That is, the expected lifetime of the top quark is sufficiently short 
to prevent long distance effects (e.g. fragmentation) from
affecting the $t\bar{t}$ spin configurations,
which are determined by the short distance dynamics of QCD at production
\cite{Bigi,Bigi2,Jezabek2,Dalitz,Dalitz2}.

The observation of spin correlation in the decay products of $t\bar{t}$ 
systems is interesting for several reasons. 
First, it provides a probe of a quark that is almost free of confinement
effects. 
Second, since the lifetime of the top quark is
proportional to the Kobayashi-Maskawa matrix element
$|V_{tb}|^2$, an observation of spin correlation would yield information
about the lower limit on $|V_{tb}|$, without 
assuming that there are three generations of 
quark families \cite{Stelzer}. Finally, many scenarios beyond
the standard model \cite{Hill,Eichten,Holdom,Lee} predict different production and decay
dynamics of the top quark, any of which could affect the observed
spin correlation.
 
In the decay of a polarized top quark, charged leptons or
quarks of weak isospin $-\frac{1}{2}$ are most sensitive to the initial polarization.
Their angular distribution in  the rest frame of the top quark
is given by $1+\cos\theta$, where $\theta$ is the angle between
the polarization direction and the line of flight of the 
charged lepton or down-type quark. Due to the experimental difficulties
of identifying jets initiated by
a down-type quark, we only consider
top-quark events in dilepton channels, i.e., where
both $W$ bosons in an event decay leptonically ($e\nu$ or $\mu\nu$).
The advantages associated with using these channels
are that: (1) objects sensitive to the polarization
of the top quark are clearly identified,
(2) background is small compared to the 
lepton+jets channels, and (3)
there are fewer ambiguities associated with assigning
objects observed in the detector to their originating quarks.
The disadvantages are that the number of events in the dilepton channels
is \mbox{small},
and that it is necessary to reconstruct two neutrinos 
in an event whose combined transverse momenta gives rise
to the observed transverse momentum imbalance in the event.

At $\sqrt{s}=1.8$ TeV, 90\% of the top quark pairs arise
from $q\bar{q}$ annihilation, and, for unpolarized incident particles,
the produced $t$ and $\bar{t}$ are also expected to be unpolarized.
However, their spins are expected to have strong correlation \cite{Stelzer,Barger} 
event by event and point along the same axis in the $t\bar{t}$ rest frame\cite{Mahlon}.
In an optimized spin quantization basis called the ``off-diagonal''
basis, contributions from opposite spin projections for
top quark pairs arising from $q\bar{q}$ annihilations are suppressed 
at the tree-level 
\cite{Mahlon} and only like spin configurations survive. 
This spin quantization basis can be specified using
the velocity  $\beta^*$ and the scattering
angle $\theta^*$ of the top quark with respect to
the center-of-mass frame of the  incoming partons.
The direction of the off-diagonal
basis forms an angle $\psi$ with respect to the $p\bar{p}$ beam axis that
is given by \cite{Mahlon,Shadmi}:
\begin{equation}
\tan\psi=\frac{\beta^{*2}\sin\theta^*\cos\theta^*}{1-\beta^{*2}\sin^2\theta^*}. 
\end{equation}
This particular choice of basis is optimal in the sense that
top quarks produced from $q\bar{q}$ will have their spins fully aligned
along this basis.
In the limit of top quark production at rest ($\beta^*=0$),
the $t$ quark and the $\bar{t}$ quark will have their spins pointing in the same
direction along $\psi=0$.

Defining $\theta_+$ as the angle between one of the charged
leptons and the axis of quantization in the rest frame of 
its parent top quark, and 
similarly defining $\theta_-$ for the other
charged lepton, the 
spin correlation can be expressed as \cite{Mahlon,Goldstein}:
\begin{equation}
\frac{1}{\sigma}\frac{d^2\sigma}{d(\cos\theta_+) d(\cos\theta_-)}
=\frac{1+\kappa\cos\theta_+\cdot\cos\theta_-}{4}, 
\label{eq:diffcross}
\end{equation}
where the correlation coefficient $\kappa$ 
describes the degree of correlation present prior
to imposition of selection criteria or effects of
detector resolutions. For $t\bar{t}$ production at
the Tevatron, the SM predicts $\kappa=0.88$\cite{mythesis}. In
the off-diagonal basis, the correlation coefficient
for $q\bar{q}\rightarrow t\bar{t}$ is $\kappa=1$. When
	$gg\rightarrow t\bar{t}$ is included at $\sqrt{s}=1.8$ TeV,
the correlation is reduced to $\kappa=0.88$. 
The distribution
is symmetric with respect to the exchange of $\theta_+$ and
$\theta_-$, and it is  therefore not necessary to identify
the electric charge of the leptons. The physical meaning of 
$\kappa$ in any spin quantization basis corresponds to
the fractional difference between the number in which the top-quark spins
are aligned and the number of events in which they have 
opposite directions.

The events used in this analysis are identical to those used
to extract the mass of the top quark in our dilepton sample \cite{top_recent}. 
They were recorded using the \dzero detector
\cite{D0_detector}, which consists of a non-magnetic tracking system
including a transition radiation detector (TRD), a 
liquid-argon/uranium calorimeter segmented in depth into
several electromagnetic (EM) and hadronic \mbox{layers}, and an outer toroidal
muon spectrometer. 
The final sample consists of three $e\mu$ events, two $ee$ events, and
one $\mu\mu$ event, with expected backgrounds of $0.21\pm 0.16$,
$0.47\pm 0.09$, and $0.73\pm 0.25$ events, respectively \cite{top_recent}.

To study the distribution in $(\cos\theta_+,\cos\theta_-)$,
we must deduce the momenta of the two neutrinos. 
The weighting scheme we use is the previously-developed neutrino weighting 
method \cite{top_recent}.
In this method, each neutrino rapidity is selected from a range of values
following a distribution consistent with the decay kinematics in $t\bar{t}$ events.
We assume the $t\bar{t}$ dilepton
decay hypothesis, and the constraints that
$m(l_1\nu_1)=m(l_2\nu_2)=m_W$
and $m(l_1\nu_1 b_1)=m(l_2\nu_2 b_2)=m_t$.
The problem can be solved   by providing a specific input mass
$m_t$ that we assume to be $m_t=175$ GeV. 
We then solve for the neutrino momentum vectors, obtaining up to  four solutions,
and assign a weight to each solution to characterize how likely
it is to represent $t\bar{t}$ production. 
A weight is assigned to each solution based on the extent
to which the sum of transverse momentum components $\sum p_k(\nu\nu)$ $(k=x,y)$
   	of the two neutrinos  
in the solution agrees with the  measured missing transverse momentum component
${\not\!\!E_k}$ $(k=x,y)$
in the event. A Gaussian distribution with a width of 4 GeV is 
assumed for each component
of the ${\not\!\!E_k}$ \cite{top_recent}. The weight is calculated as:
\begin{equation}
w^\nu=\prod_{k=x,y}\exp\left[-\frac{({\not\!\!E_k}-p_k(\nu\nu))^2}
{2\sigma^2}\right]. 
\end{equation}
The physical objects in the events are smeared
to take into consideration the finite resolution of the detector,
and we consider both possible pairings of the two charged leptons
with the two jets assigned to $b$ quarks. The presence of a third
jet is also taken into consideration \cite{top_recent}.

For each solution, we can then
boost the decay products into the rest frame of
the original top quarks and calculate the relevant decay
angles $(\cos\theta_+, \cos\theta_-)$. The event fitter
returns many such solutions for an event, and the goal is to 
deduce the original value of
$(\cos\theta_+, \cos\theta_-)$ from the reconstructed
distributions. 

The differential cross section depends on the product
$\xi=\cos\theta_+\cdot\cos\theta_-$.
We define an  asymmetry $\mathcal{A}$ for all solutions in an event as \cite{Bigi,Dalitz,Mahlon}:
\begin{equation}
{\rm \cal{A}}=\frac{1}{\sigma}\left(\int_{0}^{1} \frac{d\sigma}{d\xi} d\xi-\int_{-1}^{0} 
\frac{d\sigma}{d\xi} d\xi\right).
\end{equation}
For perfect resolution and acceptance, ${\rm \cal{A}}$ is expected
to be $\kappa/4$.

\begin{figure}
\begin{center}
\epsfig{file=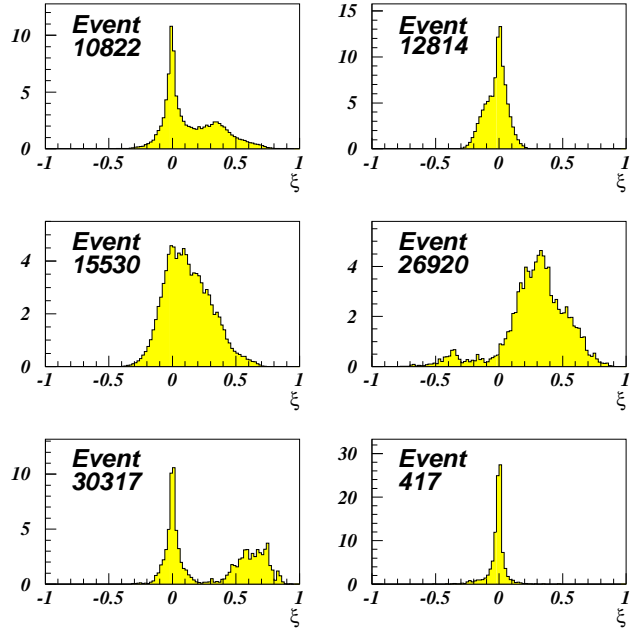,width=3.8in}
\end{center}
\caption{Distribution of $\xi$ for the 6 dilepton events.}
\label{fig:6events}
\end{figure}

Since the event fitter returns solutions with
assigned weights and there is no ``unique''
solution, we sum the weights for all the solutions to populate
the distribution $\xi$, which is shown in Fig. \ref{fig:6events} 
for the 6 events.
The values of $\mathcal{A}$ are listed in Table \ref{tab:6events}.

\begin{table}
\caption{ Asymmetry values for the 6 dilepton events at \dzero.}
\label{tab:6events}
\begin{tabular}{lcr}
Event Number & Event type & $\mathcal{A}$ \\ \tableline
10822 & $ee$ & 0.34 \\
12814 & $e\mu$  & -0.16 \\
15530 & $\mu\mu$ & 0.50 \\
26920 & $e\mu$  & 0.85 \\
30317 & $ee$ & 0.52 \\
417 & $e\mu$ & -0.19 \\ \tableline
$<\mathcal{A}>$ & & $0.31\pm 0.22$
\end{tabular}
\end{table}

 Monte Carlo event
generators such as {\sc herwig} \cite{HERWIG} and {\sc pythia} \cite{PYTHIA},
in their current implementation, do not take proper account of 
spin correlation in $t\bar{t}$ production, and the
two top quarks in an event are made to decay independently of each other,
i.e. $\kappa=0$ is assumed.
To  include the effects of spin correlation, $t\bar{t}$
events from the {\sc pythia} event generator
are sampled at the generator level
with the weight $(1+\kappa\xi)$, where
$\xi$ is calculated from information at the generator level.
We have checked this method against a Monte Carlo containing a 
fully correlated matrix element (where $\kappa=1$ for
$t\bar{t}$ events initiated from $q\bar{q}$ annihilation) and found
the two methods are equivalent~\cite{mythesis}.

To estimate the sensitivity of our method, we created 1500
ensembles of 6 events consisting of appropriate fractions
of $t\bar{t}$ signal and background.
From Monte Carlo studies, we expect ${\rm \mathcal{A}}=0.207\pm 0.006$ for
full spin correlation ($\kappa=1$) when all detector and
background effects are included, while ${\rm\mathcal{A}}=0.25$ for
perfectly reconstructed events without any background.
The statistical uncertainty on our measurements is estimated
to be 0.20 from these ensemble studies.
Similar tests were performed for ensembles of 6 events without 
spin correlation ($\kappa=0$), and  we find an expected ${\rm\mathcal{A}}=0.115\pm 0.005$,
while ideally ${\rm\mathcal{A}}=0$. 
	The main cause for loss of sensitivity
is the incorrect pairing of the 
lepton with the jet.
 This produces a strong bias in ${\rm\mathcal{A}}$~\cite{mythesis}.
From the Monte Carlo samples generated with values of $\kappa$ between
$-1$ and 1, we find  a linear relationship between ${\cal A}$
and $\kappa$: ${\cal A} = 0.112+0.088\kappa$.

We obtain ${\rm\mathcal{A}}=0.31\pm 0.22$ from
our data, which translates into $\kappa=2.3\pm 2.5$, 
assuming that a linear relationship between ${\rm\mathcal{A}}$ and
$\kappa$ also holds beyond $-1\leq\kappa\leq1$, 
though the values $|\kappa|>1$  are not physical.

Systematic uncertainties are negligible compared to the statistical
uncertainty in our result. Varying the top quark mass by 5 GeV results in a
shift in $\mathcal{A}$ of 0.01.
There has been no theoretical calculation of effects of gluon radiation on the 
spin correlation of the top quarks. However, these effects were studied
for spin-uncorrelated events (i.e. $\kappa=0$) by including
gluon radiation in the {\sc pythia} event
generator. This results in a shift in  $\mathcal{A}$  of
$0.0065\pm 0.0063$, where the error is due
to finite Monte Carlo statistics. The asymmetry distribution
expected from background is similar to that for spin-uncorrelated
$t\bar{t}$ events, and its impact is small.

To maximize the physical information
present in the data, 
the full two-dimensional phase space of $(\cos\theta_+,\cos\theta_-)$
is used in a two-dimensional binned likelihood analysis. The phase
space is split into a $3\times 3$ grid,
each side of which spans 1/3 of the range of $\cos\theta_+$ and $\cos\theta_-$.
The nine bins are populated for data with weights $(w_1,...,w_9)$
from the event fitter, with  
the distribution of weights for each event normalized to unity. 
Similar distributions
are made for the generated Monte Carlo events using different
values of $\kappa$ for $t\bar{t}$ signal and an appropriate
admixture of background. Comparisons of data with Monte Carlo
are used to extract $\kappa$.

Because
an event populates each bin with fractional probability, a simple 
likelihood assuming a Poisson distribution may not be appropriate.
Moreover, since
the weights for each event satisfy the normalization
condition $\sum_i w_i = 1$, only eight out of the nine weights are
independent, and there are correlations among the weights 
in any given event. 

To find eight independent variables, the covariance matrix 
$C_{ij}={\rm cov}(w_i,w_j),(i,j=1,\ldots,8)$ is 
calculated from the Monte Carlo events for a given spin correlation $\kappa$
and background, and diagonalized using a matrix
$A$, such that $A^{-1}CA$ has only diagonal elements. The new independent variables (i.e. diagonalized \mbox{weights})
are found by applying this transformation matrix 
to the weights, $V=A^{-1} W$, where $W=(w_1,\ldots,w_8)^T$ and 
$V=(v_1,\ldots,v_8)^T$. The distributions 
$f_i\:(i=1,\ldots,8)$ of the new variables $v_i$ are used to define the likelihood
\begin{equation}
{\rm \cal{L}}(\kappa) = \prod_i^{N}\prod_{j=1}^8 f_j(v_{ij};\kappa),
\end{equation}
where $v_{ij}$ are the new variables for $i$th event and
$N$ is the number of events. 
By explicitly constructing the likelihood, we do not have to
make any assumptions about the underlying distributions 
of the weights.

\begin{figure}
\begin{center}
\epsfig{file=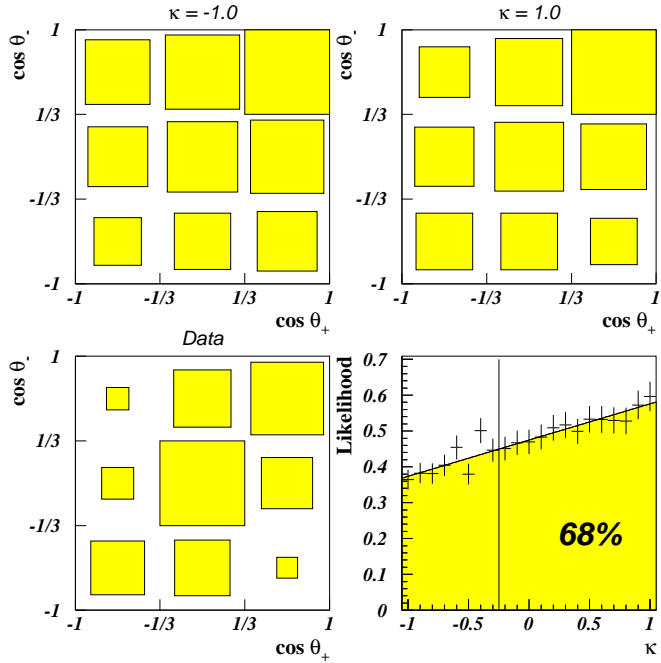,width=3.8in}
\end{center}
\caption{Plots of probability density for  $t\bar{t}$ events
in the dilepton channels in $(\cos\theta_+,\cos\theta_-)$ phase space.
Top left: Monte Carlo events with $\kappa=-1$; top right: Monte Carlo
events with $\kappa=+1$; bottom left: our data; and
bottom right: the likelihood as a function of $\kappa$ showing
the 68\% confidence limit of $\kappa>-0.25$. The box area
is proportional to the summed weights in the bin.}
\label{fig:final}
\end{figure}

The result is shown in Fig. \ref{fig:final}. The probability densities for the
Monte Carlo generator at $\kappa=-1$ and $\kappa=1$ are shown
for comparison. From the dependence of the likelihood on $\kappa$,
we can set a 68\% confidence interval at $\kappa>-0.25$, based on the line fit, in 
agreement with the SM prediction of $\kappa=0.88$.

In conclusion, we have presented a search for 
spin correlation effects in the production of $t\bar{t}$ pairs in
$p\bar{p}$ collisions at $\sqrt{s}=1.8\TeV$, where 
the dominant production mechanism is expected to be 
the annihilation of incident $q\bar{q}$ states. This analysis makes use of
the fact that there exists an optimal spin
quantization basis for the produced top quarks,
and that the charged leptons 
from top-quark decays are most sensitive to the polarization
of the top quark. From this analysis, we conclude
that $\kappa>-0.25$ at the 68\% confidence level, which is compatible with
correlation of spins expected on the basis of
the standard model.

We thank the staffs at Fermilab and at collaborating institutions 
for contributions to this work, and acknowledge support from the 
Department of Energy and National Science Foundation (USA),  
Commissariat  \` a L'Energie Atomique and
CNRS/Institut National de Physique Nucl\'eaire et 
de Physique des Particules (France), 
Ministry for Science and Technology and Ministry for Atomic 
   Energy (Russia),
CAPES and CNPq (Brazil),
Departments of Atomic Energy and Science and Education (India),
Colciencias (Colombia),
CONACyT (Mexico),
Ministry of Education and KOSEF (Korea),
CONICET and UBACyT (Argentina),
A.P. Sloan Foundation,
and the Humboldt Foundation.


\begin{references}

\bibitem{top_papers}
CDF Collaboration, F. Abe \etal, \Journal{\PRL}{74}{2626}{1995};
\dzero Collaboration, S. Abachi \etal, \Journal{\PRL}{74}{2632}{1995}.

\bibitem{top_recent}
\dzero Collaboration, S. Abachi \etal, \Journal{\PRL}{79}{1197}{1997};
\dzero Collaboration, S. Abachi \etal, \Journal{\PRL}{79}{1203}{1997};
\dzero Collaboration, B. Abbott \etal, \Journal{\PRD}{58}{052001}{1998};
\dzero Collaboration, B. Abbott \etal, \Journal{\PRL}{80}{2063}{1998};
\dzero Collaboration, B. Abbott \etal, \Journal{\PRD}{60}{012001}{1999};
\dzero Collaboration, B. Abbott \etal, \Journal{\PRD}{60}{052001}{1999};
\dzero Collaboration, B. Abbott \etal, \Journal{\PRL}{83}{1908}{1999}.

\bibitem{CDF}
CDF Collaboration, F. Abe \etal, \Journal{\PRL}{79}{1992}{1997};
CDF Collaboration, F. Abe \etal, \Journal{\PRL}{79}{3585}{1997};
CDF Collaboration, F. Abe \etal, \Journal{\PRL}{80}{2767}{1998};
CDF Collaboration, F. Abe \etal, \Journal{\PRL}{80}{2773}{1998};
CDF Collaboration, F. Abe \etal, \Journal{\PRL}{80}{2779}{1998};
CDF Collaboration, F. Abe \etal, \Journal{\PRL}{82}{271}{1999};
CDF Collaboration, F. Abe \etal, \Journal{\PRD}{59}{092001}{1999}.


\bibitem{Jezabek} M. Je\.{z}abek, \Journal{\NPBproc}{37}{197}{1994}.

\bibitem{PDG} Particle Data Group, C. Caso \etal, \Journal{\EPC}{3}{1}{1998}. 

\bibitem{Peskin} A.F.  Falk and M.E. Peskin, 
	\Journal{\PRD}{49}{3320}{1994}.

\bibitem{Bigi} I. Bigi, \Journal{\PLB}{175}{233}{1986}.

\bibitem{Bigi2} I. Bigi, Y. Dokshitzer, V. Khoze, J. K\"{u}hn, and P.
Zerwas, \Journal{\PLB}{181}{157}{1986}.

\bibitem{Jezabek2} M. Je\.{z}abek and J. K\"{u}hn, \Journal{\PLB}{329}{317}{1994}.

\bibitem{Dalitz} R.H. Dalitz and G.R. Goldstein, \Journal{\PRD}{45}{1531}{1992}.

\bibitem{Dalitz2} R.H. Dalitz and G.R. Goldstein, \Journal{Int. Jour. Mod. Phys.}{A9}{635}{1994}.

\bibitem{Stelzer} T. Stelzer and S. Willenbrock, \Journal{\PLB}{374}{169}{1996}.

\bibitem{Hill} C. Hill and S. Parke, \Journal{\PRD}{49}{4454}{1994}.

\bibitem{Eichten} E. Eichten and K. Lane, \Journal{\PLB}{327}{129}{1994}.

\bibitem{Holdom} B. Holdom and T. Torma \Journal{\PRD}{60}{114010}{1999}.

\bibitem{Lee} Kang Young Lee \etal, \Journal{\PRD}{60}{093002}{1999}.

\bibitem{Barger} V. Barger, J. Ohnemus, and R.J.N. Phillips, \Journal{Int. J. Mod. Phys.}{A4}{617}{1989}.

\bibitem{Mahlon} G. Mahlon and S. Parke, \Journal{\PRD}{53}{4886}{1996};
G. Mahlon and S. Parke, \Journal{\PLB}{411}{173}{1997}.

\bibitem{Goldstein} G.R. Goldstein, in {\em Spin 96: Proceedings of the 12th International
	Symposium on High Energy Spin Physics, Amsterdam, 1996}, edited by C.W. deJager
	(World Scientific, 1997), p. 328.

\bibitem{Shadmi} S. Parke and Y. Shadmi, \Journal{\PLB}{387}{199}{1996}.


\bibitem{mythesis} S. Choi, Ph.D. dissertation, Seoul National University, 1999
(unpublished) \\http://www-d0.fnal.gov/results/publi\-cations\_talks/thesis/choi/thesis.ps.



\bibitem{D0_detector} \dzero Collaboration, S. Abachi \etal, 
	\Journal{\NIM\  in Phys. Res. A}{338}{185}{1994}.

\bibitem{HERWIG} G. Marchesini \etal, \Journal{\CPC}{67}{465}{1992}.

\bibitem{PYTHIA} T. Sj\"ostrand, \Journal{\CPC}{82}{74}{1994}.


\end{references}
\end{document}